\documentclass[12pt,aps,epsfig,amssymb,tighten]{revtex4}
\usepackage{epsfig}
\begin{document}
\title{Randomly accelerated particle in a box: mean absorption time
for partially absorbing and inelastic boundaries}
\author{Stanislav N. Kotsev and Theodore W. Burkhardt\\
Department of Physics, Temple University, Philadelphia, PA 19122}


\date{\today}
\begin{abstract}
Consider a particle which is randomly accelerated by Gaussian white noise on the
line segment $0<x<1$ and is absorbed as soon as it reaches $x=0$ or $x=1$. The mean
absorption time $T(x,v)$, where $x$ and $v$ denote the initial position and
velocity, was calculated exactly by Masoliver and Porr\`a in 1995. We consider a
more general boundary condition. On arriving at either boundary, the particle is
absorbed with probability $1-p$ and reflected with probability $p$. The reflections
are inelastic, with coefficient of restitution $r$. With exact analytical and
numerical methods and simulations, we study the mean absorption time as a function
of $p$ and $r$.
\end{abstract}
\maketitle \vskip 2cm PACS 02.50.Ey, 05.40.-a, 45.70.-n

\newpage
\section{Introduction}
\label{sec:intro} Consider a randomly accelerated particle which
moves on the half line $x>0$ with an absorbing boundary at $x=0$.
The motion is governed by
\begin{equation}{d^2x\over
dt^2}=\eta
(t)\;,\quad\langle\eta(t)\eta(t')\rangle=
2\delta(t-t')\;,\label{eqmo}
\end{equation}
where $\eta(t)$ is uncorrelated white noise with zero mean. The probability
$Q(x,v;t)$ that a particle with initial position $x$ and initial velocity $v$ has
not yet been absorbed after a time $t$ has been studied by McKean \cite{mck} and
others \cite{goldman,ys,twb93}. For a particle initially at $x=0$ with velocity
$v>0$, the probability decays as
\begin{equation}
Q(0,v;t)\sim \left({v^2\over t}\right)^{1/4}\;,\quad t\to\infty\;.\label{Qabs}
\end{equation}

Following Cornell, Swift, and Bray \cite{csb}, we consider a more general boundary
condition. On arriving at the boundary the particle is absorbed with probability
$1-p$ and reflected with probability $p$. The velocities of the particle just after
and before reflection are related by $v_f=-rv_i$. Here $r$ is the coefficient of
restitution, and $r=1$ and $r<1$ correspond to elastic and inelastic collisions,
respectively.

As physical motivation, we note that temporal or spatial irregularities at the
boundaries of a system or a statistical capture process may give rise to partial
absorption. The properties of a particle which is subject to a random force and
collides inelastically are of interest in connection with driven granular media.

For the partially absorbing, inelastic boundary condition the probability that the
particle has not yet been absorbed decays as
\begin{equation}
Q(0,v;t)\sim \left({v^2\over t}\right)^{\phi(p,r)}\;,\quad t\to\infty\;.\label{Qpr}
\end{equation}
Burkhardt \cite{twb2000} and De Smedt {\it et al.} \cite{dsgl} showed that the
persistence exponent $\phi(p,r)$ is non-universal, depending on $p$ and $r$
according to the relation
\begin{equation}
pr^{2\phi}=2\sin\left({1-4\phi\over 6}\pi\right)\;.\label{expopr}
\end{equation}
For $p=0$, $\phi(0,r)={1\over 4}$, so Eqs. (\ref{Qabs}) and (\ref{Qpr}) are
consistent. Solving Eq. (\ref{expopr}) for $\phi$ graphically, one finds a unique
solution $0<\phi(p,r)<{1\over 4}$ in the physical region $0<r<1$, $0<p<1$,
representing a decay slower than in Eq. (\ref{Qabs}).

 The decay in Eq. (\ref{Qpr}) is so slow that the mean absorption time
\begin{equation} T(0,v)=\int_0^\infty dt\thinspace t\left[-{\partial
Q(0,v;t)\over\partial t}\right]=\int_0^\infty dt\thinspace Q(0,v;t)\label{meantime}
\end{equation}
is infinite, even for $p=0$. On the other hand, the mean absorption time $T^*(0,v)$,
averaged over a long but finite time $t^*$ instead of an infinite time, is a
well-defined quantity, which varies as
\begin{equation}
T^*(0,v)\sim t^*\left(v^2\over t^*\right)^{\phi(p,r)}\;,\quad t^*>>v^2\label{Tstar}
\end{equation}
and diverges in the limit $t^*\to\infty$.

If the particle moves on the finite line $0<x<1$ and is absorbed the first time it
reaches either $x=0$ or $x=1$, the probability that it has not yet been absorbed
after a time $t$ decays as $e^{-\lambda t}$ for long times \cite{twb97}, much more
rapidly than the power law (\ref{Qabs}) for a single absorbing boundary. The mean
absorption time $T(x,v)$ for a particle with initial postion $x$ and initial
velocity $v$ is well defined. Masoliver and Porr\`a \cite{mp} calculated this
quantity exactly, by solving the inhomogeneous Fokker-Planck equation
\begin{equation}
\left(v{\partial\over\partial x}+{\partial^2\over \partial
v^2}\right)T(x,v)=-1\;,\label{fp}
\end{equation}
with the boundary conditions
\begin{eqnarray}
&&T(x,v)=T(1-x,-v)\;,\label{refsym}\\
&&T(0,v)=0\;,\quad v<0\;,\label{absorbingbc}
\end{eqnarray}
corresponding to reflection symmetry and the immediate absorption
of a particle with initial conditions $x=0$, $v<0$. Their result,
which is rederived in the Appendix, is given by
\begin{eqnarray}
&&T_1(x,v)=T_1(1-x,-v)={1\over 3}(\pi v)^{1/2}\int_0^{1-x} dy\thinspace
{e^{-v^3/18y}\over y^{1/2}}I_{-1/6}\left({v^3\over 18 y}\right)
+{2^{-7/3}3^{5/6}\over\Gamma(2/3)^2}\nonumber\\
&&\quad\times\int_x^1 dy\thinspace {e^{-v^3/9(y-x)}\over
(y-x)^{2/3}}[y(1-y)]^{-1/6}\thinspace\left[_2F_1(1,-{2\over 3};{5\over
6};1-y)-\thinspace_2F_1(1,-{2\over 3};{5\over 6};y)\right]\label{mp}
\end{eqnarray}
for $v>0$, where $_2F_1(a,b;c;z)$ is a standard hypergeometric function \cite{gr}.
The subscript $1$ of $T_1(x,v)$ is a reminder that the particle is absorbed the
first time it reaches either boundary. For use below we note the asymptotic forms
\begin{equation}
T_1(0,v)\sim\left\{\begin{array}{l}{\displaystyle {3\over
\pi^{1/2}}\thinspace v^{1/2}} \;,\quad v\searrow 0\\
\\{\displaystyle {1\over v}},\quad v\to\infty
\end{array}\right.\label{asymp}
\end{equation}
The second of these forms is intuitively obvious, corresponding to ballistic
propagation from $x=0$ to $x=1$ in the time $v^{-1}$.

In this paper we study the mean absorption time of a randomly-accelerated particle
moving on the finite line $0<x<1$ for the more general boundary condition described
above: absorption at the boundary with probability $1-p$, reflection with
probability $p$ and coefficient of restitution $r$. For this boundary the mean
absorption time also satisfies the Fokker-Planck equation (\ref{fp}) with reflection
symmetry (\ref{refsym}), but the absorbing boundary condition (\ref{absorbingbc}) is
replaced by
\begin{equation}
T(0,-v)=pT(0,rv)\;,\quad v>0\;.\label{partialabsorbingbc}
\end{equation}

The absorbing boundary condition (\ref{absorbingbc}) considered by Masoliver and
Porr\`a is unusual in that $T(0,v)$ is specified for $v<0$ but not $v>0$. That the
Fokker-Planck equation with this boundary condition and with reflection symmetry
(\ref{refsym}) is a well-posed boundary value problem was known \cite{fichera} long
before the exact solution of Masoliver and Porr\`a. We are unaware of a similar
proof for the more general boundary condition (\ref{partialabsorbingbc}). In this
paper we show how to solve the more general case with an extension of the approach
of Masoliver and Porr\`a.

\section{Integral equation for $T(x,v)$}

In the Appendix we show that the Fokker-Planck equation (\ref{fp}) with reflection
symmetry (\ref{refsym}) has the exact Green's function solution
\begin{equation}
T(x,-v)=T_1(x,-v)+\int_0^\infty du\thinspace u\thinspace
G(x,v,u)T(0,-u)\;,\label{gf1}
\end{equation}
for $v>0$, where $T_1$ is the same function as in Eq. (\ref{mp}),
\begin{eqnarray}
&&G(x,v,u)={v^{1/2}u^{1/2}\over 3x}\thinspace
e^{-(v^3+u^3)/9x}I_{-1/3}\left({2v^{3/2}u^{3/2}\over 9x}\right)\nonumber\\
&&\qquad\qquad\qquad -{1\over 3^{1/3}\Gamma({2\over 3})}\int_0^x dy\thinspace
{e^{-v^3/9(x-y)}\over (x-y)^{2/3}}\left[R(y,u)-R(1-y,u)\right]\;,\label{gf2}
\end{eqnarray}
\begin{equation} R(y,u)={1\over 3^{5/6}\Gamma({1\over 3})\Gamma({5\over
6})}\thinspace {u^{1/2}e^{-u^3/9y}\over y^{7/6}(1-y)^{1/6}}\; _1F_1 \left(-{1\over
6},{5\over 6},{u^3(1-y)\over 9y}\right)\;,\label{gf3}
\end{equation}
and $_1F_1(a,b,z)$ is a standard confluent hypergeometric function
\cite{gr}.

To calculate $T(x,-v)$ from Eq. (\ref{gf1}), one must first
determine the unknown function $T(0,-u)$ on the right-hand side.
Setting $x=1$ in Eqs. (\ref{gf1})-(\ref{gf3}) and substituting
$T(1,-v)=T(0,v)$ and $T(0,-u)=pT(0,ru)$, which follow from Eqs.
(\ref{refsym}) and (\ref{partialabsorbingbc}), leads to the
integral equation
\begin{equation}
T(0,v)=T_1(0,v)+p\int_0^\infty du\thinspace uG(1,v,u)T(0,ru)\;,\quad v>0\;,\label{ieq1}\\
\end{equation}
where
\begin{equation}
G(1,v,u)={1\over 6\pi}v^{1/2}u^{1/2}\thinspace e^{-(v^3+u^3)/9}\left[{9\over
v^3+u^3}+6\; _1F_2\left(1;{5\over 6},{7\over 6};{v^3u^3\over
81}\right)\right]\;,\label{ieq2}
\end{equation}
and $_1F_2(a;b,c;z)$ is a generalized hypergeometric function \cite{gr}. Our
analytical and numerical predictions for the mean absorption time are derived
directly from Eqs. (\ref{ieq1}) and (\ref{ieq2}). The same symmetric kernel
$G(1,v,u)=G(1,u,v)$ plays a central role in Refs. \cite{bfg,bk}, in determining the
equilibrium distribution function $P(x,v)$ of a randomly accelerated particle moving
between inelastic walls at $x=0$ and $x=1$ with which it collides inelastically.

As discussed in \cite{bk}, the quantity $G(1,v,u)$ generalizes McKean's result for
the velocity distribution at first return \cite{velocitydistribution} from the
half-line to the finite line. The probability that a randomly accelerated particle,
which leaves $x=0$ with velocity $v>0$, arrives with speed between $u$ and $u+du$
the next time it reaches either $x=0$ or $x=1$ is given by $uG(1,v,u)du$, where the
first and second terms of Eq. (\ref{ieq2}) correspond to arrival at $x=0$ and $x=1$,
and where
\begin{equation}
\int_0^\infty du\thinspace u G(1,v,u)=1\;.\label{normalization}
\end{equation}
Integral equation (\ref{ieq1}) follows immediately from this interpretation. The
first, second, $\dots$ terms in the iterative solution of the integral equation
represent the mean time to reach the boundary for the first time, the mean time
between the first and second boundary collisions, \dots, weighted with a factor $p$
for each reflection.

In Eq. (\ref{ieq1}) the asymptotic form of $T(0,v)$ for small and large $v$ is
determined by the first and second terms, respectively, of the kernel (\ref{ieq2}).
For $p>0$ the particle is not always absorbed the first time it reaches the
boundary, which implies $T(0,v)>T_1(0,v)$. For small $v$ we look for a solution with
the asymptotic form
\begin{equation}
T(0,v)\sim v^\gamma\;,\quad v\searrow 0\;,\label{asy1}
\end{equation}
with exponent $\gamma(p,r)$ smaller than the value $\gamma={1\over 2}$ for $p=0$ in
Eq. (\ref{asymp}). Substituting Eq. (\ref{asy1}) in Eq. (\ref{ieq1}), we find that
the asymptotic form is consistent with the integral equation if
\begin{equation}
pr^\gamma=2\sin\left({1-2\gamma\over 6}\thinspace\pi\right)\;.\label{asy2}
\end{equation}
Solving Eq. (\ref{asy2}) for $\gamma$ graphically, one finds a unique solution
satisfying $0<\gamma(p,r)<{1\over 2}$ in the physical region $0<r<1$, $0<p<1$.
Comparing Eqs. (\ref{expopr}) and (\ref{asy2}), one sees that
$\gamma(p,r)=2\phi(p,r)$. Thus, the exponents of $v$ in the results (\ref{expopr}),
(\ref{Tstar}) for the half line and (\ref{asy1}), (\ref{asy2}) for the finite line
are the same. This suggests that the small $v$ behavior (\ref{asy1}), (\ref{asy2})
is due to repeated collisions with the same boundary.

For large $v$, $T(0,v)\approx v^{-1}+T(1,v)$, corresponding to ballistic propagation
from $x=0$ to $x=1$ in the time $v^{-1}$. Combining this relation with Eqs.
(\ref{refsym}) and (\ref{partialabsorbingbc}) yields
\begin{equation}
T(0,v)\approx v^{-1}+ pT(0,rv)\;,\quad v\to\infty\;.\label{asy3}
\end{equation}
Equation (\ref{asy3}) also follows from integral equation (\ref{ieq1}) on using the
asymptotic forms (\ref{asymp}) for large $v$ and $_1F_2\left(1;{5\over 6},{7\over
6};z\right)\approx {1\over 6}\pi^{1/2}z^{-1/4}\exp(2z^{1/2})$ for large $z$. For
$p<1$, the mean absorption time is expected to decrease to zero in the large $v$
limit. Iteration of Eq. (\ref{asy3}) and/or substitution of the ansatz
$T(0,v)\approx Bv^{-\nu}$ in Eq. (\ref{asy3}) leads to the asymptotic forms
\begin{equation}
T(0,v)\approx\left\{\begin{array}{l}\displaystyle {r\over r-p}\thinspace
v^{-1}\;,\\\displaystyle Bv^{-\ln p/\ln
r}\;,\end{array}\right.\quad\begin{array}{l}p<r\leq 1\;,\\r<p<1\;,\end{array}
\label{asy4}
\end{equation}
for $v\to\infty$, with $p<1$.

The large $v$ behavior for $p=1$, $r<1$, corresponding to inelastic reflection with
zero absorption probability, requires special attention and is considered in Section
V-C.

\section{Numerical Solution of the Integral Equation}
To interpolate between the asymptotic forms of $T(0,v)$ for small and large $v$,
given in the preceding Section, we solved integral equation (\ref{ieq1})
numerically. Changing the integration variable from $u$ to $ru$, we solved the
resulting equation,
\begin{equation}
T(0,v)=T_1(0,v)+{p\over r^2}\int_0^\infty du\thinspace uG(1,v,r^{-1}u)T(0,u)\;,\quad
v>0\label{ieq3}\;,
\end{equation}
by iteration.

To evaluate the integral in Eq. (\ref{ieq3}) numerically, we changed the integration
variable from $u$ to
\begin{equation}
\begin{array}{l}\displaystyle z={u^3\over 9r^3}\;,\\\displaystyle
w=\left({u^3\over 9 r^3}\right)^{-\alpha}\;,
\end{array}\quad\begin{array}{l}0<u<B\;,\\B<u<\infty\;,\end{array}
\label{variables}
\end{equation}
where $\alpha$ is a positive constant, and then used Simpson's rule, exact for
polynomials of degree 3, with equally spaced mesh points in $z$ and $w$. To ensure
good accuracy for small $z$ (small $u$), we subtracted off the leading asymptotic
form of the integrand, fit to a power law, and then integrated it analytically. In
the region where both $v$ and $u$ are small, where the first term in the kernel
(\ref{ieq2}) diverges, the integrand was fit to $(v^3+u^3)^{-1}$ times a power law
in $z$ and then integrated analytically. The parameters $B$ and $\alpha$ and the
mesh size were chosen to give good convergence with about 100 mesh points in each of
the intervals $0<B<u$ and $B<u<\infty$. Starting with $T_1(0,v)$ as the first
approximation to $T(0,v)$, one typically needs about 10 iterations for convergence.

\section{Simulations}

In our simulations the motion of the particle is governed by the difference
equations
\begin{eqnarray}
x_{n+1}&=&x_{n}+v_n\Delta_{n+1}+\left({\Delta_{n+1}^3
\over6}\right)^{1/2}(s_{n+1}+\sqrt3
r_{n+1})\;,\label{xstep}\\
v_{n+1}&=&v_{n}+\left(2\Delta_{n+1}\right)^{1/2}r_{n+1}\;.\label{vstep}
\end{eqnarray}
Here $x_n$ and $v_n$ are the position and velocity at time $t_n$, and
$\Delta_{n+1}=t_{n+1}-t_n$. The quantities $r_n$ and $s_n$ are independent Gaussian
random numbers such that
\begin{equation}
\langle r_n \rangle=\langle s_n \rangle=0\;,\quad\langle r_n^2 \rangle=\langle s_n^2
\rangle=1\;.\label{gauss}
\end{equation}

As discussed in \cite{bibu}, this algorithm generates trajectories which are
consistent with the exact probability distribution $P(x,v,t)$ of a randomly
accelerated particle in free space, i.e. in the absence of boundaries. In free space
there is no time step error in the algorithm. The time step $\Delta_{n+1}$ is
arbitrary. However, close to the boundaries trajectories are not generated with the
correct probability, because the free space distribution $P(x,v,t)$ includes
trajectories which wander outside the interval $0<x<1$ and return during the time
$t$. As in \cite{bibu}, we make the time step smaller near the boundaries to exclude
these spurious trajectories.

Using the algorithm (\ref{xstep})-(\ref{gauss}) with large time steps away from the
boundaries, instead of a conventional algorithm with a constant time step, enables
us to simulate the particle for much longer times.

For a particle in free space with position and velocity $x_n$, $v_n$ at time $t_n$,
the coordinate $x_{n+1}$ at a time $\Delta_{n+1}$ is distributed according to a
Gaussian function \cite{bibu}, with a maximum at $x_n+v_n\Delta_{n+1}$ and root-mean
square width $\left({2\over 3}\Delta_{n+1}^3\right)^{1/2}$. Choosing $\Delta_{n+1}$
so that
\begin{eqnarray}
&&x_{n} + v_n\Delta_{n+1}-c\Delta_{n+1}^{3/2}>0\;,\label{step1}\\
&&x_{n} + v_n\Delta_{n+1}+c\Delta_{n+1}^{3/2}<1\;,\label{step2}
\end{eqnarray}
where the constant $c$ is about 5 or larger, ensures that the Gaussian distribution
lies well within the interval $0<x<1$.

The largest $\Delta_{n+1}$ consistent with inequality (\ref{step1}) is given by
$D(x_n,v_n)$, where
\begin{equation}
x+ vD(x,v)-cD(x,v)^{3/2}=0\;.\label{Ddef}
\end{equation}
To avoid solving Eq. (\ref{Ddef}), cubic in $D^{1/2}$, at each step of the
algorithm, one may use the approximation
\begin{equation} D(x,v)\approx\left\{\begin{array}{l}\displaystyle\left({x\over
c}\right)^{2/3}+\left({v\over c}\right)^2\;,\\
\displaystyle\left[\left({c\over x}\right)^2-{v\over
x}\right]^{-1}\;,\end{array}\right.\quad\begin{array}{l}v>0\;,\\v<0\;,
\end{array}\label{Dapprox}
\end{equation}
which is asymptotically exact for $v\to 0$ and $v\to\pm\infty$ and accurate to
better than $13\%$ for $-\infty<v<\infty$.

The most efficient time step to use in the algorithm is the largest $\Delta_{n+1}$
consistent with {\em both} inequalities (\ref{step1}) and (\ref{step2}). Using a
smaller $\Delta_{n+1}$ slows the simulation without improving the accuracy. Since
the inequality (\ref{step2}) follows from (\ref{step1}) on making the reflection
symmetric substitution $v_n\to -v_n$, $x_n\to 1-x_n$, the optimal time step is
\begin{equation}
\Delta_{n+1}={\rm Min}[D(x_n,v_n),D(1-x_n,-v_n)]+\delta\;,\label{beststep}
\end{equation}
where Min denotes the smaller of the two quantities inside the square brackets. As
in \cite{bibu} a small minimum time step $\delta$ has been included in Eq.
(\ref{beststep}). Without it, the step size decreases to zero as the particle
approaches the boundary, and, as in Zeno's paradox, the particle never gets there.
After inclusion of $\delta$, the particle not only reaches the boundary but jumps
slightly pass it. We kept the overshoot small by choosing a sufficiently large value
for the parameter $c$ in Eqs. (\ref{step1})-(\ref{Dapprox}).

In the case $p=1$, $r<r_c$ of highly inelastic collisions with absorption
probability zero, the speed of the particle becomes extremely small after many
boundary collisions, and great care is required to simulate the behavior reliably.
According to Eqs. (\ref{vstep}) and (\ref{beststep}), the root-mean-square velocity
change at each time step has the minimum average value
\begin{equation}
\Delta v=(2\delta)^{1/2}\;.\label{deltav}
\end{equation}
After each boundary collision we set $\Delta v$ equal to 1/500 of the velocity just
after the collision. This $\Delta v$ is the smallest velocity the algorithm can
reliably handle. The corresponding value of $\delta$, given by Eq. (\ref{deltav}),
is then used until the next boundary collision.

The results for the mean time $T(0,v)$ reported in Section V are based on $10^4$
independent particle trajectories for each value of the initial velocity $v$.

\section{Results}
\subsection{Partially absorbing, elastic boundaries}

In this subsection we present our various results for $p<1$, $r=1$. In this regime
the particle is absorbed at the boundary with probability $1-p$ and reflected {\em
elastically} with probability $p$. This is the ``partial survival" model, studied on
the half line in Refs. \cite{csb,sb,twb2000,dsgl}.

For $p<1$, $r=1$ the mean absorption time has the asymptotic forms (\ref{asy1}),
(\ref{asy2}), and (\ref{asy4}) for small and large $v$, respectively. As the
reflection probability $p$ increases from 0 to 1, $\gamma(p)$ decreases from the
Masoliver-Porra result $\gamma(0)={1\over 2}$ in Eq. (\ref{asymp}) for absorption on
first arrival at the boundary, to the value $\gamma(1)=0$, corresponding to no
absorption at all, or $T(0,v)=\infty$.

\begin{figure}[h]
\begin{center}
\includegraphics*[width=0.7\textwidth]{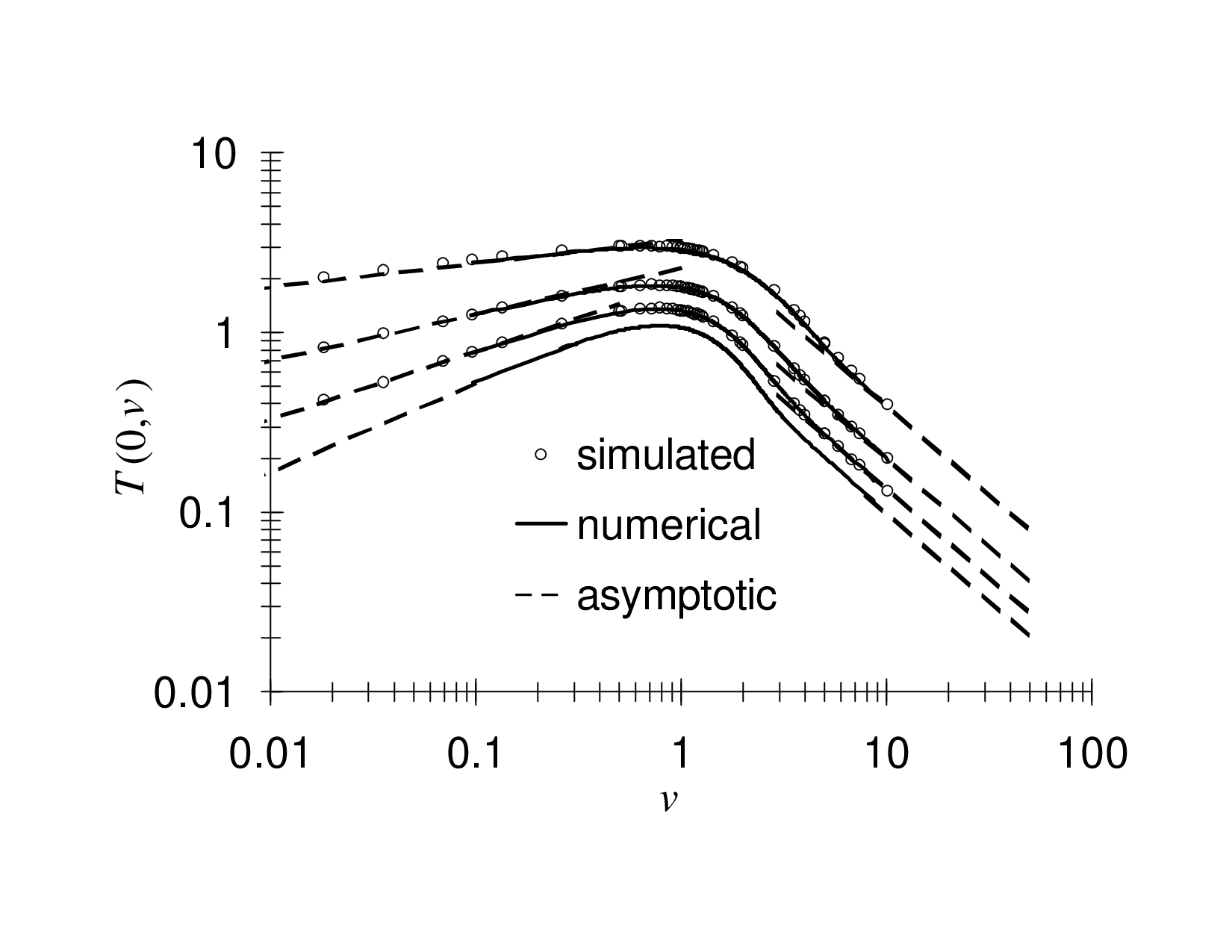}\\*
\end{center}
\caption{Mean absorption time versus initial velocity for $r=1$ and, from top to
bottom, $p=0.75$, $0.50$, $0.25$, and $0$. The solid line for $p=0$ is the exact
result of Masoliver and Porr\`a. The other solid lines are the numerical solutions
of the integral equation (\ref{ieq1}). The points are the results of our
simulations. The dashed lines show the exact asymptotic forms (\ref{asy1}),
(\ref{asy2}) for small $v$ and (\ref{asy4}) for large $v$. The proportionality
constant in Eq. (\ref{asy1}) was chosen to fit the simulation data.}\label{pfig}
\end{figure}

In Fig. \ref{pfig} our results for the mean absorption time $T(0,v)$ for partially
absorbing, elastic boundary conditions as a function of the initial velocity $v$ are
compared for $r=1$ and $p=0.75,\;0.5,\;0.25,\;0$. The exact asymptotic forms
(\ref{asy1}) and (\ref{asy2}) for small $v$ and (\ref{asy4}) for large $v$, the
numerical solution of Eq. (\ref{ieq1}), and the simulation results are clearly in
excellent agreement.

For fixed $v$ the mean absorption time in Fig. \ref{pfig} increases with increasing
reflection probability $p$, as expected. For fixed $p$ the mean absorption time does
not vary monotonically with $v$ but has an absolute maximum at an intermediate
velocity. For much larger initial velocities the particle is absorbed more rapidly
since it bounces back and forth between the boundaries, colliding at a rapid rate.
For much smaller initial velocities the particle is absorbed more rapidly because it
collides repeatedly with the boundary where it starts.

\subsection{Partially absorbing, inelastic boundaries}
Here we present our results for $p<1$, $r<1$. In this regime the particle is
absorbed at the boundary with probability $1-p$ and reflected {\em inelastically}
with probability $p$ and coefficient of restitution $r$. Results for the half-line
geometry with this boundary condition are given in Ref. \cite{twb2000}.

\begin{figure}[h]
\begin{center}
\includegraphics*[width=0.7\textwidth]{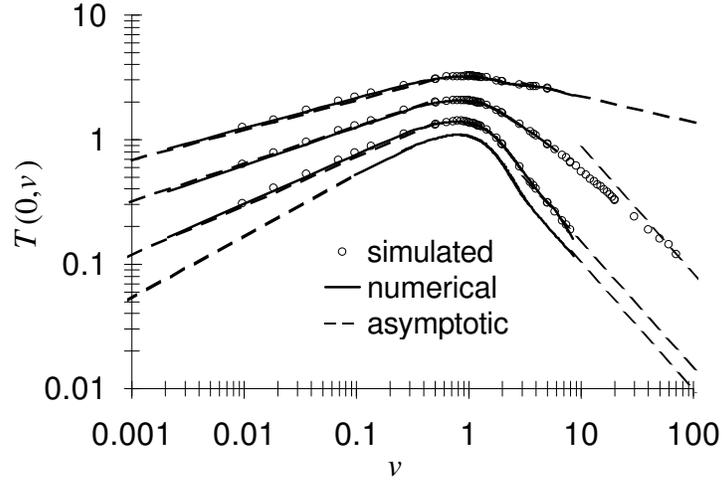}\\*
\end{center}
\caption{Mean absorption time versus initial velocity for, from top to bottom,
$(p,r)=(0.75,0.25)$, $(0.501,0.5)$, $(0.25,0.75)$, $(0,r)$. The bottom curve is the
exact result (\ref{mp}) of Masoliver and Porr\`a. The other solid lines are the
numerical solutions of the integral equation (\ref{ieq1}). The points are the
results of our simulations. The dashed lines show the exact asymptotic forms
(\ref{asy1}), (\ref{asy2}) for small $v$ and (\ref{asy4}) for large $v$. The
proportionality constant in Eq. (\ref{asy1}) was chosen to fit the simulation data.
Note the slow convergence to the asymptotic form for large $v$ for (0.501,0.5) in
the crossover region $p\approx r$.} \label{requalpfig}
\end{figure}

In Fig. \ref{requalpfig} our results for the mean absorption time $T(0,v)$ as a
function of the initial velocity $v$ are compared for $(p,r)=(0.75,0.25)$,
$(0.501,0.5)$, and $(0.25,0.75)$. Again the exact asymptotic forms (\ref{asy1}) and
(\ref{asy2}) for small $v$ and (\ref{asy4}) for large $v$, the numerical solution of
Eq. (\ref{ieq1}), and the simulation results are in excellent agreement. Note that
the large $v$ asymptotic forms (\ref{asy4}) for $p<r<1$ and $r<p<1$ are different
and that the data test both. Corrections to the large $v$ asymptotic form in the
crossover regime $p\approx r$ lead to conspicuously slower convergence for
$(p,r)=(0.501,0.5)$.

\subsection{Non-absorbing, inelastic boundaries}
In this subsection we consider boundaries with $p=1$, $r<1$. At the boundary the
particle is reflected inelastically with probability 1 and absorbed with probability
zero. This is the case of interest in connection with driven granular matter.
Cornell {\it et al.} \cite{csb} predicted that for $r<r_c$, where
\begin{equation}
r_c=e^{-\pi/\sqrt 3}=0.163\dots, \;\label{rc}
\end{equation}
the particle undergoes ``inelastic collapse," making an infinite number of boundary
collisions in a finite time, coming to rest, and remaining there.

The prediction that the inelastic collisions localize the particle at the boundary
for $r<r_c$ was questioned by Florencio {\it et al.} \cite{fbb} and Anton \cite{la}
on the basis of simulations.

The equilibrium distribution function $P(x,v)$ of a particle moving on the finite
line between two inelastic boundaries was studied by Burkhardt, Franklin, and
Gawronski \cite{bfg} for $r>r_c$ and Burkhardt and Kotsev \cite{bk} for $r<r_c$ with
exact analytical and numerical calculations and simulations, similar to this paper.
According to \cite{bk}, $P(x,v)$ varies smoothly and analytically with $r$
throughout the interval $0<r<1$ and does not collapse onto the boundaries. For
$r<r_c$ the equilibrium boundary collision rate is infinite, but the collisions do
not localize the particle at the boundary (see footnote \cite{collrate}).

In the case $p=1$, $r<1$ iterating integral equation (\ref{ieq1}) adds the mean time
to reach the boundary for the first time, the mean time between the first and second
boundary collisions, etc., taking the change in speed in each collision into
account. Thus the solution $T(0,v)$ represents the mean time $T(0,v)$ for an
infinite number of boundary collisions.

For $p=1$, $r<r_c$, $T(0,v)$ has the asymptotic forms (\ref{asy1}), (\ref{asy2}) for
small $v$. As the coefficient of restitution $r$ increases from 0 to 1, $\gamma(r)$
decreases from the Masoliver-Porra result $\gamma(0)={1\over 2}$ in Eq.
(\ref{asymp}) to the value $\gamma(r_c)=0$, confirming the result of Cornell {\it et
al.} for $r_c$ in Eq. (\ref{rc}). The negative, unphysical value of $\gamma$ for
$r>r_c$ signals the breakdown of the solution with finite $T(0,v)$. For $r>r_c$ the
mean time for an infinite number of collisions is infinite.

In contrast with the results for $p<1$ given in Eq. (\ref{asy4}), in the case $p=1$,
$r<1$, the mean time $T(0,v)$ does not vanish in the limit $v\to\infty$. This may be
understood by reverse iteration of the large $v$ recurrence relation (\ref{asy3}),
which yields
\begin{equation}
T\left(0,r^{-n}v\right)\approx \left(r+r^2+\dots+r^n\right)\thinspace
v^{-1}+T(0,v)\;.\label{Tinfty}
\end{equation}
This result holds for any finite velocity $v$ large enough so that the time to
travel between the two boundaries is accurately given by the ballistic time
$v^{-1}$. The first $n$ terms on the right-hand side give the time a particle with
initial velocity $r^{-n} v$ takes to reach velocity $v$ and sum to
$r(1-r)^{-1}v^{-1}$ in the limit $n\to\infty$.

According to Eq. (\ref{Tinfty}),
$\lim_{n\to\infty}T\left(0,r^{-n}v\right)=T(0,\infty)$ is finite and non-vanishing
for $r<r_c$. The most general asymptotic form for large $v$ consistent with these
two properties and with the recurrence relation (\ref{asy3}) is
\begin{equation} T(0,v)\approx\left\{\begin{array}{l}\infty\;,\\
\displaystyle f_{\rm per}(\ln v)-{r\over 1-r}\thinspace
v^{-1}\;,\end{array}\right.\quad\begin{array}{l} p=1\;,\;\;r>r_c\;,
\\p=1\;,\;\;r<r_c\;,\end{array} \label{asy5}
\end{equation}
where $f_{\rm per}(\ln v)$ is a periodic function of $\ln v$ with period $|\ln r|$.

The periodic term in Eq. (\ref{asy5}) came as a surprise to us. The large $v$
recurrence relation (\ref{asy3}) implies $T(0,v)>T(0,rv)$ for $p=1$, which suggests,
but does not guarantee, that $T(0,v)$ is a monotonically increasing function of $v$.
The asymptotic form (\ref{asy5}) for $r<r_c$ with the periodic term is entirely
consistent with $T(0,v)>T(0,rv)$ and with Eq. (\ref{Tinfty}). The periodicity
already appears in the following crude approximation: For $0<v<v^*$ define
$T(0,v)=Av^\gamma$, in accordance with Eq. (\ref{asy1}), and then use the large $v$
recurrence relation, Eq. (\ref{asy3}) with $p=1$, to obtain $T(0,v)$ for $v>v^*$.

\begin{figure}[h]
\begin{center}
\includegraphics*[width=0.7\textwidth]{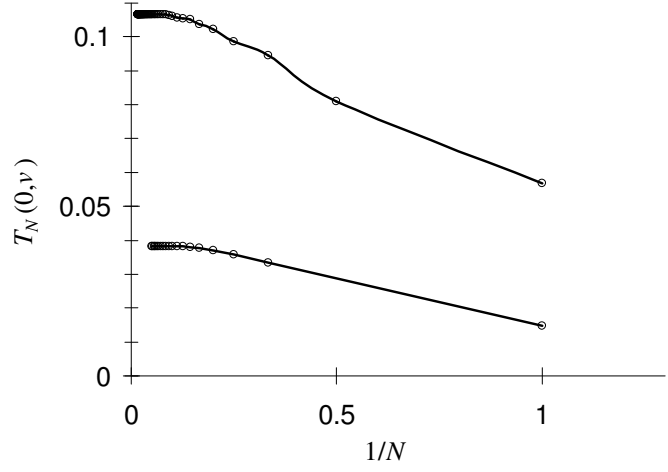}\\*
\end{center}
\caption{Mean time $T_N(0,v)$ for $N$ boundary collisions versus $N^{-1}$ for
$r=0.01$ and, from top to bottom, $v=$0.001 and 0.0001. The mean time $T(0,v)$ for
an infinite number of collisions is estimated by extrapolating to the vertical axis.
This yields $T(0,0.001)=0.108$ and $T(0,0.0001)=0.039$.}\label{extrapfig}
\end{figure}

To determine $T(0,v)$ with simulations, we measured the mean time $T_N(0,v)$ for $N$
boundary collisions, plotted it versus $N^{-1}$, and then extrapolated to
$N\to\infty$. This is shown for $r=0.1$ and $v=0.001$ and $0.0001$ in Fig.
\ref{extrapfig}.

\begin{figure}[h]
\begin{center}
\includegraphics*[width=0.7\textwidth]{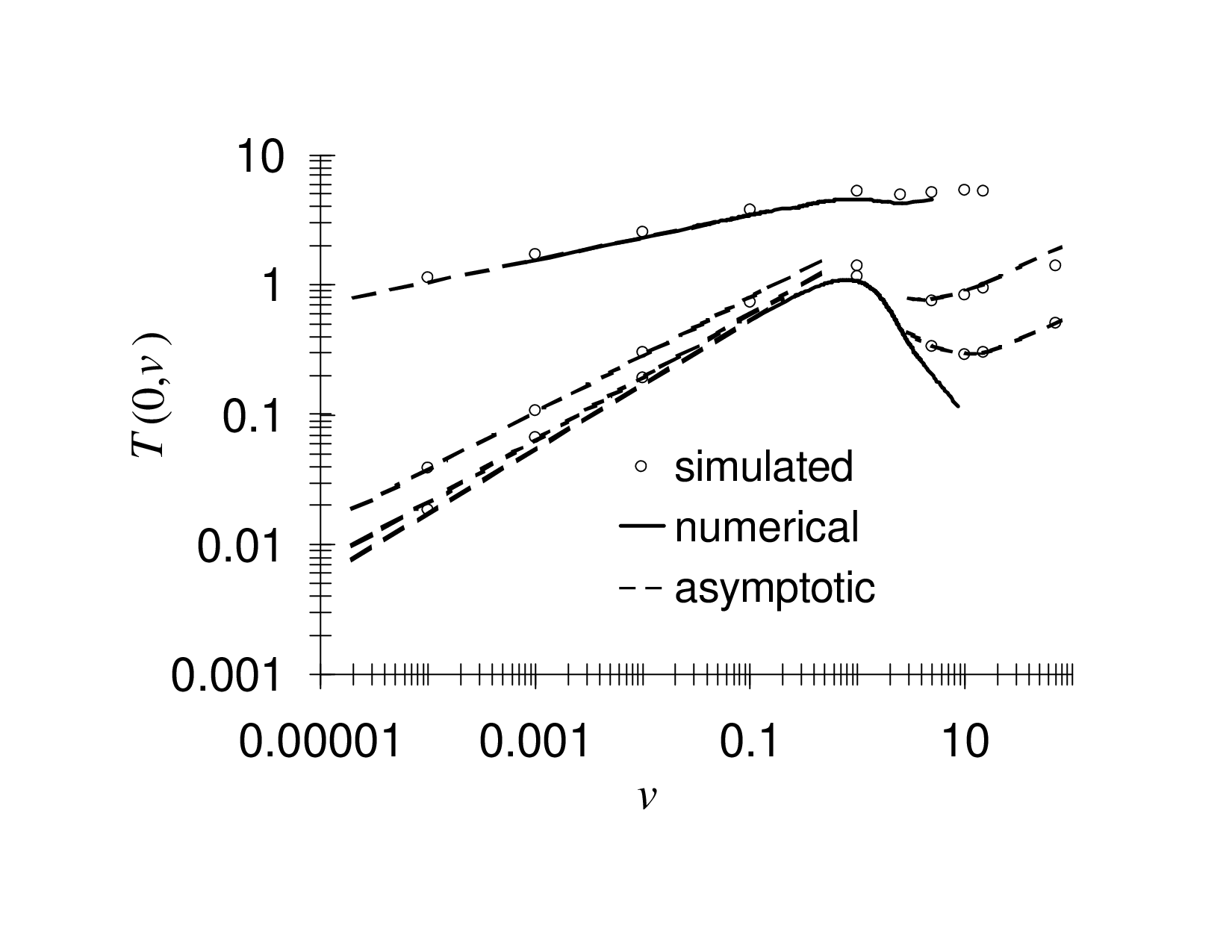}\\*
\end{center}
\caption{Mean time for an infinite number of boundary collisions for $p=1$ and, from
top to bottom, $r=0.1$, $0.01$, $0.001$. The upper curve is the numerical solution
of the integral equation (\ref{ieq1}) for $r=0.1$. The points are the results of our
simulations, obtained by extrapolation, as in Fig. \ref{extrapfig}. The bottom curve
(no simulations) is the exact Masoliver-Porr\`a result (\ref{mp}). The dashed lines
on the left show the exact asymptotic form (\ref{asy1}), (\ref{asy2}) for small $v$.
The proportionality constant in Eq. (\ref{asy1}) was chosen to fit the simulation
data. For large $v$, approximately periodic behavior in $\ln v$, as in Eq.
(\ref{asy5}), is expected. The dashed lines for large $v$ in Fig. \ref{rfig} were
calculated from the dashed lines for small $v$ using recurrence relation
(\ref{asy3}).}\label{rfig}
\end{figure}

In Fig. \ref{rfig} our results for the mean time $T(0,v)$ for an infinite number of
boundary collisions are shown as a function of $v$ for $ p=1$ and $r=0.1$, $0.01$,
and $0.001$. For fixed $v$ the mean time increases with increasing $r$, as expected.

Solving integral equation (\ref{ieq1}) reliably becomes increasingly difficult as
$r$ decreases, and in Fig. \ref{rfig} we only show the numerical solution for
$r=0.1$. The simulation data are in good agreement with this numerical solution,
with the asymptotic form (\ref{asy1}), (\ref{asy2}) for small $v$, and with the
onset of periodic behavior in accordance with Eq. (\ref{asy5}). The dashed lines for
large $v$ were calculated from the dashed lines for small $v$ using recurrence
relation (\ref{asy3}).

\section{Concluding Remarks}
Masoliver and Porr\`a calculated the mean absorption time $T(x,v)$ of a randomly
accelerated particle on the finite line $0<x<1$, assuming that it is absorbed the
first time it reaches either $x=0$ or $x=1$. We have considered a more general
boundary condition, parametrized by the reflection probability $p$ and the
coefficient of restitution $r$. We derived an integral equation which determines
$T(x,v)$ and used it to obtain the exact asymptotic form of $T(0,v)$ for large and
small $v$ and numerical results for intermediate $v$. The asymptotic forms and
numerical results are in excellent agreement with our computer simulations of the
randomly accelerated particle.

The case $p=1$, $r<1$ corresponds to a single particle moving between walls with
which it collides inelastically. This simple and fundamental system is of interest
in connection with properties of driven granular media. We find that the mean time
for an infinite number of boundary collisions is finite for $r<r_c$ and infinite for
$r>r_c$, as predicted by Cornell {\it et al.} \cite{csb}. The prediction that the
infinite sequence of collisions leads to inelastic collapse, with localization of
the particle at the boundary, has been questioned on the basis of simulations
\cite{fbb,la} and an analysis \cite{bk} of the equilibrium distribution function
$P(x,v)$ using the same approach as in this paper. The results for the mean time
reported here are largely independent of the inelastic collapse controversy. Both
the integral equation and the simulation procedure sum the mean time, collision by
collision, without directly addressing the question of localization at the boundary.

\acknowledgements We thank Jerrold Franklin for many stimulating discussions. TWB
also gratefully acknowledges discussions and correspondence with Lucian Anton and
Alan Bray.

\appendix
\section{Solution of Eq. (\ref{fp})}

With the substitution $\tilde{T}(x,v)=T(x,-v)$, Eq. (\ref{fp}) takes the form
\begin{equation}
\left(v{\partial\over\partial x}-{\partial^2\over \partial
v^2}\right)\tilde{T}(x,v)=1\;.\label{fp2}
\end{equation}
This differential equation is the same as Eq. (3) of Ref. \cite{bfg}, except for the
inhomogeneous term $1$ on the right-hand side. We follow Ref. \cite{bfg} very
closely in solving it.

First we introduce the Laplace transform
\begin{equation}
\tilde{Q}(s,v)=\int_0^\infty dx\thinspace e^{-sx}\tilde{T}(x,v)\;,
\end{equation}
which according to Eq. (\ref{fp2}) satisfies
\begin{equation}
\left(sv-{\partial^2\over \partial v^2}\right)\tilde{Q}(s,v)={1\over s}
+v\tilde{T}(0,v)\;.\label{LTfp2}
\end{equation}
This the same as Eq. (A2) in \cite{bfg}, except for the extra term $s^{-1}$ on the
right-hand side. As in \cite{bfg}, we solve Eq. (\ref{LTfp2}) in terms of Airy
functions and invert the Laplace transform, with the help of the Faltung theorem.
This yields
\begin{eqnarray}
\tilde{T}(x,v)&=&{1\over3}(\pi v)^{1/2}\int_0^x dy\thinspace
y^{-1/2}e^{-v^3/18y}I_{-1/6}\left({v^3\over 18y}\right)\nonumber\\&\ &+{v^{1/2}\over
3x}\int_0^\infty du\thinspace
u^{3/2}e^{-(v^3+u^3)/9x}I_{-1/3}\left({2v^{3/2}u^{3/2}\over 9x}\right)\tilde{T}(0,u)\nonumber\\
&\ &-{1\over 3^{1/3}\Gamma({2\over 3})}\int_0^xdy\thinspace {e^{-v^3/9(x-y)}\over
(x-y)^{2/3}}\thinspace {\partial\tilde{T}(y,0)\over\partial v}\label{Agf1}\;,\quad
v>0\;,
\end{eqnarray}
which is the same as Eq. (6) of \cite{bfg}, except for an additional term (the first
term) on the right-hand side.

To express the unknown function $\partial\tilde{T}(y,0)/\partial v$ on the
right-hand side of Eq. (\ref{Agf1}) in terms of the other unknown $\tilde{T}(0,u)$,
we first take the limit $v=0$ in Eq. (\ref{Agf1}), which yields
\begin{eqnarray}
\tilde{T}(x,0)&=&{1\over 3^{1/3}\Gamma({2\over 3})}\Big[\thinspace{3^{2/3}\over
2}\Gamma(\textstyle{1\over 3})x^{2/3}+x^{-2/3}\displaystyle\int_0^\infty
du\thinspace u e^{-u^3/9x}\thinspace\tilde{T}(0,u)\nonumber\\&\ &-\int_0^x {dy\over
(x-y)^{2/3}}\thinspace {\partial\tilde{T}(y,0)\over\partial
v}\thinspace\Big]\label{Agf2}\;.
\end{eqnarray}
From reflection symmetry $\tilde{T}(x,0)-\tilde{T}(1-x,0)=0$, and
$\partial\tilde{T}(y,0)/\partial v=-\partial\tilde{T}(1-y,0)/\partial v$.
Substituting Eq. (\ref{Agf2}) in the first of these relations and making use of the
second, we obtain
\begin{equation}
\int_0^1{dy\over|x-y|^{2/3}}\thinspace{\partial\tilde{T}(y,0)\over\partial v}=
\int_0^\infty du\thinspace u\left[{e^{-u^3/9x}\over x^{2/3}}-{e^{-u^3/9(1-x)}\over
(1-x)^{2/3}}\right]\left(\textstyle{1\over
2}u^2+\tilde{T}(0,u)\right)\;,\label{Agf3}
\end{equation}
where we have used
\begin{equation}
\int_0^\infty du\thinspace u^3{e^{-u^3/9x}\over
x^{2/3}}=3^{2/3}\Gamma(\textstyle{1\over 3})x^{2/3}\;.
\end{equation}
Equation (\ref{Agf3}) may be regarded as an integral equation for
$\partial\tilde{T}(y,0)/\partial v$. The solution, derived in \cite{bfg} using the
approach of \cite{mp,ps}, is
\begin{equation}
{\partial\tilde{T}(y,0)\over\partial v}=\int_0^\infty du\thinspace
u\left[R(y,u)-R(1-y,u)\right]\left(\textstyle{1\over 2}u^2+\tilde{T}(0,u)\right)\;,
\label{Agf4}
\end{equation}
where the function $R(y,u)$ is given in Eq. (\ref{gf3}).

Substituting Eq. (\ref{Agf4}) in (\ref{Agf1}) yields
\begin{eqnarray}
\tilde{T}(x,v)&=&{1\over3}(\pi v)^{1/2}\int_0^x dy\thinspace
y^{-1/2}e^{-v^3/18y}I_{-1/6}\left({v^3\over 18y}\right)\nonumber\\&\ & -{1\over
2\cdot 3^{1/3}\Gamma({2\over 3})}\int_0^xdy\thinspace {e^{-v^3/9(x-y)}\over
(x-y)^{2/3}}\thinspace\int_0^\infty du\thinspace
u^3\left[R(y,u)-R(1-y,u)\right]\nonumber\\&\ &+\int_0^\infty du\thinspace
u\thinspace G(x,v,u)\tilde{T}(0,u)\;,\label{Agf5}
\end{eqnarray}
where $G(x,v,u)$ is given in Eqs. (\ref{gf2}). Replacing $\tilde{T}(x,v)$ by
$T(x,-v)$ in Eq. (\ref{Agf5}) and evaluating the integral $\int_0^\infty
du\thinspace u^3[\dots]$, we obtain the Green's function solution for $T(x,-v)$ in
Eqs. (\ref{gf1})-(\ref{gf3}) and reproduce the Masoliver-Porr\`a result for
$T_1(x,v)$ in Eq. (\ref{mp}).

\newpage
\vskip3ex

\newpage
\vskip3ex
\vskip3ex

\newpage
\vskip3ex
\vskip3ex

\newpage
\vskip3ex
\vskip3ex


\begin{thebibliography}{03}
\bibitem{collrate}  This result, which may seem counter-intuitive, has the following
mathematical origin: The equilibrium distribution $P(x,v)$ obtained in \cite{bfg,bk}
varies asymptotically as $P(0,v)\sim v^{-\beta(r)}$ for $v\searrow 0$, where the
exponent $\beta(r)$ increases monotonically from $0$ to ${5\over 2}$ as $r$
decreases from 1 to 0, with $\beta(r_c)=2$. Thus, the equilibrium collision rate
$I=\int_{0}^{\infty}dv v P(0,v)$ diverges at the lower limit for $r<r_c$. This
result only holds for the continuous time dynamics (\ref{eqmo}). For discrete
dynamics with a nonvanishing time step, the number of collisions in a finite time is
necessarily finite.
\end{thebibliography}
\end{document}